\documentstyle[psfig]{mn}

\newif\ifAMStwofonts

\ifoldfss
  \ifCUPmtlplainloaded \else
    \NewTextAlphabet{textbfit} {cmbxti10} {}
    \NewTextAlphabet{textbfss} {cmssbx10} {}
    \NewMathAlphabet{mathbfit} {cmbxti10} {} 
    \NewMathAlphabet{mathbfss} {cmssbx10} {} 
  \fi
  \ifAMStwofonts
    \ifCUPmtlplainloaded \else
      \NewSymbolFont{upmath} {eurm10}
      \NewSymbolFont{AMSa} {msam10}
      \NewMathSymbol{\upi}     {0}{upmath}{19}
      \NewMathSymbol{\umu}     {0}{upmath}{16}
      \NewMathSymbol{\upartial}{0}{upmath}{40}
      \NewMathSymbol{\leqslant}{3}{AMSa}{36}
      \NewMathSymbol{\geqslant}{3}{AMSa}{3E}

    \fi
  \fi
\fi 

\ifnfssone
  \newmathalphabet{\mathit}
  \addtoversion{normal}{\mathit}{cmr}{m}{it}
  \addtoversion{bold}{\mathit}{cmr}{bx}{it}
  \newmathalphabet{\mathbfit} 
  \addtoversion{normal}{\mathbfit}{cmr}{bx}{it}
  \addtoversion{bold}{\mathbfit}{cmr}{bx}{it}
  \newmathalphabet{\mathbfss} 
  \addtoversion{normal}{\mathbfss}{cmss}{bx}{n}
  \addtoversion{bold}{\mathbfss}{cmss}{bx}{n}
  \ifAMStwofonts
    \ifCUPmtlplainloaded \else
      \UseAMStwoboldmath
      \makeatletter
      \new@mathgroup\upmath@group
      \define@mathgroup\mv@normal\upmath@group{eur}{m}{n}
      \define@mathgroup\mv@bold\upmath@group{eur}{b}{n}
      \edef\UPM{\hexnumber\upmath@group}
      \new@mathgroup\amsa@group
      \define@mathgroup\mv@normal\amsa@group{msa}{m}{n}
      \define@mathgroup\mv@bold\amsa@group{msa}{m}{n}
      \edef\AMSa{\hexnumber\amsa@group}
      \makeatother
      \mathchardef\upi="0\UPM19
      \mathchardef\umu="0\UPM16
      \mathchardef\upartial="0\UPM40
      \mathchardef\leqslant="3\AMSa36
      \mathchardef\geqslant="3\AMSa3E
    \fi
  \fi
\fi 

\ifnfsstwo
  \DeclareMathAlphabet{\mathbfit}{OT1}{cmr}{bx}{it}
  \SetMathAlphabet\mathbfit{bold}{OT1}{cmr}{bx}{it}
  \DeclareMathAlphabet{\mathbfss}{OT1}{cmss}{bx}{n}
  \SetMathAlphabet\mathbfss{bold}{OT1}{cmss}{bx}{n}
  \ifAMStwofonts
    \ifCUPmtlplainloaded \else
      \DeclareSymbolFont{UPM}{U}{eur}{m}{n}
      \SetSymbolFont{UPM}{bold}{U}{eur}{b}{n}
      \DeclareSymbolFont{AMSa}{U}{msa}{m}{n}
      \DeclareMathSymbol{\upi}{0}{UPM}{"19}
      \DeclareMathSymbol{\umu}{0}{UPM}{"16}
      \DeclareMathSymbol{\upartial}{0}{UPM}{"40}
      \DeclareMathSymbol{\leqslant}{3}{AMSa}{"36}
      \DeclareMathSymbol{\geqslant}{3}{AMSa}{"3E}
    \fi
  \fi
\fi 

\ifCUPmtlplainloaded \else
  \ifAMStwofonts \else 
    \def\upi{\pi}
    \def\umu{\mu}
    \def\upartial{\partial}
  \fi
\fi

\title{The 3.4 micron absorption feature towards three obscured 
active galactic nuclei
}
\author[Imanishi]
       {M.~Imanishi $^{1,2}$ \\
$^1$ National Astronomical Observatory, Mitaka, Tokyo 181-8588, Japan \\
$^2$ Institute for Astronomy, University of Hawaii, 
2680 Woodlawn Drive, Honolulu, Hawaii 96822, USA\\
} 
\date{} 

\pubyear{2000}

\begin{document} 

\maketitle


\begin{abstract} 

The results of 3--4 $\mu$m spectroscopy towards the nuclei of NGC 3094, 
NGC 7172, and NGC 7479 are reported.  
In ground-based 8--13 $\mu$m spectra, all the sources have strong 
absorption-like features at $\sim$10 $\mu$m, but they do not have 
detectable polycyclic aromatic hydrocarbon (PAH) emission features.  
The 3.4 $\mu$m carbonaceous dust absorption features are detected 
towards all nuclei.
NGC 3094 shows a detectable 3.3 $\mu$m PAH emission feature, while 
NGC 7172 and NGC 7479 do not.
Nuclear emission whose spectrum shows dust absorption features but 
no PAH emission features is thought to dominated by 
highly obscured active galactic nuclei (AGNs) activity.
For NGC 7172, NGC 7479, and three other such nuclei in the literature, 
we investigate the optical depth ratios between the 3.4 $\mu$m 
carbonaceous dust and 9.7 $\mu$m silicate dust absorption 
($\tau_{3.4}$/$\tau_{9.7}$).
The $\tau_{3.4}$/$\tau_{9.7}$ ratios towards three highly obscured AGNs 
with face-on host galaxies are systematically larger than the ratios 
in the Galactic diffuse interstellar medium or the ratios for 
two highly obscured AGNs with edge-on host galaxies.
We suggest that the larger ratios can be explained if the 
obscuring dust is so close to the central AGNs that 
a temperature gradient occurs in it.
If this idea is correct, our results may provide spectroscopic 
evidence for the presence of the putative ``dusty tori'' 
in the close vicinity of AGNs.

\end{abstract} 
 

\section{INTRODUCTION} 
This is the second paper in a series that investigates the 3.4 $\mu$m
carbonaceous dust absorption feature in sources that show a strong
absorption-like feature at $\sim$10 $\mu$m but no detectable
polycyclic aromatic hydrocarbon (PAH) emission features in
ground-based 8--13 $\mu$m (i.e., whole the $N$-band) spectra.
The first paper discusses NGC 5506 (Imanishi 2000).
Roche et al. (1991) performed 
extensive ground-based 8--13 $\mu$m spectroscopy of the nuclei of 
nearby active galaxies (mostly $z$ $<$ 0.05),
and classified them into the following three groups:
(1) those dominated by the family of emission features from small
molecules called PAHs (20--40 carbon atoms per molecule; Allamandola,
Tielens \& Barker 1989),
(2) those with a featureless continuum, and 
(3) those that display an absorption-like feature at $\sim$10 $\mu$m.
Each group of sources is regarded as, respectively, 
(1) those powered by star-forming activity, 
(2) those powered by unobscured active galactic nuclei (AGNs) activity, 
and 
(3) those powered by obscured AGN activity, for which the silicate dust 
absorption feature at $\sim$9.7 $\mu$m is expected to be present.
Sources in the third group, in particular those with a strong silicate 
dust absorption feature, are expected to be highly obscured AGNs, 
and thus can be used to investigate the properties of the intervening 
medium along our line-of-sight through the study of infrared absorption 
features.

However, the attribution of the strong absorption-like feature 
at $\sim$10 $\mu$m detected in many ground-based 8--13 $\mu$m spectra 
to the 9.7 $\mu$m silicate dust absorption has been 
questioned based on the {\it Infrared Space Observatory} ({\it ISO}) 
spectra (Clavel et al. 1999; Genzel et al. 1998).
Strong PAH emission at 7.7 $\mu$m from star-forming activity and 
continuum flux that increases with wavelength 
could produce an apparent strong absorption-like feature at 
$\sim$10 $\mu$m. 
One can attempt to estimate the strength of the 7.7 $\mu$m PAH emission 
based on the observed flux of the PAH emission at 8.6 $\mu$m and 
11.3 $\mu$m in ground-based 8--13 $\mu$m spectra 
(Dudley 1999) by assuming that the 7.7 $\mu$m PAH emission 
is weak in sources with no detectable 8.6 $\mu$m and 11.3 $\mu$m PAH 
emission. 
However, this estimate could sometimes be uncertain 
(Rigopoulou et al. 1999).

Spectroscopy at 3--4 $\mu$m can be a powerful method for finding 
a highly obscured AGN, 
not only because the effect of extinction is small at 3--4 
$\mu$m, but also because there exist emission and absorption features
for distinguishing between highly obscured AGN and star-forming 
activities.
If a source is powered by star-forming activity, PAH emission features
should be detected at 3.3 $\mu$m and sometimes at 3.4 $\mu$m 
(Tokunaga et al. 1991).  
If a source is powered by
obscured AGN activity, a 3.4 $\mu$m absorption feature associated with
carbonaceous dust grains (Pendleton et al. 1994) should be detected.
An advantage of a ground-based 3--4 $\mu$m spectrum over a
ground-based 8--13 $\mu$m spectrum is that there is no serious
uncertainty in determining a continuum level for the former. This is
because both the long and short wavelength sides of the features at
3.3--3.4 $\mu$m are observable in the $L$-band atmospheric window 
(2.8--4.2 $\mu$m) for a nearby ({\it z} $<$ 0.18) source.  Furthermore, 
if both (less obscured) star-forming activity and obscured AGN activity 
are associated in a galaxy, the 3.4 $\mu$m absorption feature is 
suppressed 
because the observed spectrum is dominated by emission from the less 
obscured star-forming activity and/or the emission at 3.4 $\mu$m 
by PAHs could veil the absorption feature.  
Therefore, sources with a strong 3.4 $\mu$m absorption feature that 
also lacks a PAH emission feature are the ones most likely to be 
highly obscured AGNs with little star-forming activity.  
Unfortunately, high-quality 3--4 $\mu$m spectra for sources with a 
strong absorption-like feature at
$\sim$10 $\mu$m but no detectable PAH emission features in
ground-based 8--13 $\mu$m spectra are available only for a few
objects.  Examples include NGC 1068 (Bridger, Wright \& Geballe 1994;
Imanishi et al. 1997), IRAS 08572+3915 
(Pendleton 1996; Wright et al. 1996) and NGC 5506
(Imanishi 2000).

Here, we have performed 3--4 $\mu$m spectroscopy on three such sources, 
with the aim of detecting the 3.4 $\mu$m carbonaceous dust
absorption feature.

\section{TARGETS}

\subsection{NGC 3094}

NGC 3094 ({\it z} = 0.008) is a spiral galaxy (Huang et al. 1996).
Its nuclear optical spectrum is classified as an AGN-type 
(Armus, Heckman \& Miley 1989). 
A ground-based 8--13 $\mu$m spectrum taken
with a 3$\farcs$8 aperture displays a strong absorption-like feature
at $\sim$10 $\mu$m, but no detectable PAH emission features (Roche et
al. 1991).

\subsection{NGC 7172}

NGC 7172 ({\it z} = 0.009) is an edge-on galaxy of type S0--Sa with a
clear equatorial dust lane along roughly the east-west direction
(Anupama et al. 1995).  An obscured AGN was found in the hard X-ray
(Piccinotti et al. 1982).  A ground-based 8--13 $\mu$m spectrum taken
with a 4$\farcs$2 aperture displays a strong absorption-like feature
at $\sim$10 $\mu$m, but no detectable PAH emission features (Roche et
al. 1991).

\subsection{NGC 7479}

NGC 7479 ({\it z} = 0.008) is a spiral galaxy of type SBbc (Sandage \&
Tammann 1987).  Its grand design spiral arms are clearly seen in the
Digitized Sky Survey image, and therefore the viewing direction towards
the galaxy is estimated to be far from edge-on (Ma, Peng \& Gu 1998).
Recent high-quality optical to near-infrared spectra suggest the
presence of an obscured AGN in NGC 7479 (Ho, Filippenko \& Sargent
1997; Larkin et al. 1998).  A ground-based 8--13 $\mu$m spectrum taken
with a 4$\farcs$3 aperture displays a strong absorption-like feature
at $\sim$10 $\mu$m, but no detectable PAH emission features (Roche et
al. 1991).

\section{OBSERVATION AND DATA ANALYSIS}  

The 3--4 $\mu$m spectra towards NGC 3094, NGC 7172, and NGC 7479 were 
obtained at the 3.8-m United Kingdom Infrared Telescope (UKIRT) 
on Mauna Kea, Hawaii, using the cooled grating spectrometer 
(CGS4; Mountain et al. 1990).  
An observing log is summarized in Table 1.
The sky was photometric throughout the observations.  
The detector was a 256$\times$256 InSb array.  
A 40 l mm$^{-1}$ grating with a 2 pixel wide slit (= 1$\farcs$2) was used.

Spectra were obtained towards the flux peak at $\sim$3.5 $\mu$m.  
The slit position angles were set along the east-west direction 
for NGC 3094, and along the north-south direction for NGC 7172 
and NGC 7479. 
A standard nodding technique along the slit direction with an 
amplitude of 11$\farcs$6 was employed to subtract the signal
from the sky.  
The air masses during the observations and total on-source integration 
time are summarized in Table 1.
Standard stars (Table 1) were observed with almost the same 
air masses as the target objects to correct for the 
transmission of Earth's atmosphere.  
The main features in the spectra to be investigated were 
the 3.3 $\mu$m PAH emission and the 3.4 $\mu$m
carbonaceous dust absorption features.  
Type-A rather than type-G stars were chosen as standard stars, 
in order to avoid possible small effects of stellar photospheric 
OH absorption at 3.0--3.6 $\mu$m in late-type stars 
(Smith, Sellgren \& Brooke 1993).

Standard data analysis procedures were employed using IRAF 
\footnote{
IRAF is distributed by the National Optical Astronomy Observatories, 
which are operated by the Association of Universities for Research 
in Astronomy, Inc. (AURA), under cooperative agreement with the 
National Science Foundation. 
}.  
After the obtained frames were bias subtracted, they were
divided by a flat image.  The spectra of the targets and the standard
stars were extracted using an optimal extraction algorithm.
Wavelength calibration was performed using a Krypton or Argon lamp.  
The signals of the targets were divided by those of the standard stars 
and then multiplied by the spectrum of the blackbody with a 
temperature corresponding to individual standard stars (Table 1) 
to obtain the final spectra.

As a consequence of selecting type-A stars as standard stars, the
resulting spectra are affected by strong stellar absorption features
at 3.30 $\mu$m (Pf$\delta$) and at 3.74 $\mu$m (Pf$\gamma$).  Since
all targets are nearby, the (relatively narrow) Pf emission lines are
affected by these absorption features, and therefore will not be
discussed.  When making the final spectra, the data at 3.31--3.32
$\mu$m have been removed, because the signals there are extremely
small due to the strong methane absorption by Earth's atmosphere.
Since the data at 3.45--3.46 $\mu$m and 3.49--3.50 $\mu$m for 
NGC 7172 and NGC 7479 are affected by large noise levels 
that are difficult to correct for completely,
these data points have also been removed.  Since the 3.3 $\mu$m
PAH emission and the 3.4 $\mu$m carbonaceous dust absorption features
have broad spectral profiles, their discussion is not seriously affected 
by the removal of these data points.

\section{RESULTS}

\subsection{NGC 3094}

A flux-calibrated spectrum of NGC 3094 is shown in Fig. 1.
Our data give an $L$-band magnitude of 8.4 mag (1$\farcs$2
$\times$ 5$''$ aperture), which is almost consistent to the 
photometric magnitude at $L'$ measured with a 5$''$ aperture 
(Dudley 1998; $L'$ = 129.8 mJy or 8.2 mag).

In Fig. 1, a broad absorption-like feature is recognizable at
$\sim$3.35--3.55 $\mu$m.  
We interpret the feature as the 3.4 $\mu$m carbonaceous dust absorption.
In the Galaxy, the wavelength range of the 3.4 $\mu$m carbonaceous
dust absorption feature is 3.33--3.55 $\mu$m in the rest-frame
(Pendleton et al. 1994), which corresponds to 3.36--3.58 $\mu$m at the
redshift of {\it z} = 0.008.  The data points at $>$3.58 $\mu$m are
therefore not affected by this absorption feature.  To avoid possible
contamination by the Pf$\gamma$ emission line of NGC 3094, the data
points at 3.58--3.72 $\mu$m are regarded as the continuum level on the
long wavelength side of the absorption feature.  On the short
wavelength side of the absorption feature, the data points at $<$3.26
$\mu$m are regarded as the continuum, because they are affected
neither by the possible 3.3 $\mu$m PAH emission feature (3.25--3.35
$\mu$m in the rest-frame; Tokunaga et al. 1991) nor by the Pf$\delta$
emission line.  The best linear fit to the data points at $<$3.26
$\mu$m and at 3.58--3.72 $\mu$m is adopted as the continuum line for
the 3.4 $\mu$m carbonaceous dust absorption feature.  The adopted
continuum level is displayed in Fig. 1 as a solid line.  The resulting
optical depth relative to this continuum level is $\tau_{3.4}$ $\sim$
0.04 near the absorption peak.  There may be some uncertainties in
determining the continuum level.  
We also tried to determine the continuum level
by varying the wavelength range slightly,
and the resulting values for
$\tau_{3.4}$ were within 10\% of the value derived above for any
reasonable continuum levels adopted.

In Fig. 1, a broad emission-like feature is found at $\sim$3.3 $\mu$m.
This feature is interpreted as the 3.3 $\mu$m PAH emission. 
Its rest-frame equivalent width is estimated to be 
3.5$\pm$0.2 $\times$ 10$^{-3}$ $\mu$m.

\subsection{NGC 7172}

A flux-calibrated spectrum of NGC 7172 is shown in Fig. 2.
Our data give an $L$-band magnitude of 9.1 mag (1$\farcs$2
$\times$ 5$''$ aperture), which is fainter than the values of 8.0--9.0
mag measured with 5$''-$8$''$ apertures (Sharples et al. 1984; Glass
\& Moorwood 1985; Lawrence et al. 1985; Kotilainen et al. 1992).
The remaining $L$-band flux may originate from extended stellar
emission roughly along the east-west direction, but a quantitative
discussion of the origin of the $L$-band emission is difficult due to
the large time variation of the $L$-band flux (Sharples et al. 1984).

In Fig. 2, a broad absorption-like feature is recognizable at
$\sim$3.35--3.55 $\mu$m.  
We regard the feature as the 3.4 $\mu$m carbonaceous dust absorption.
A continuum level is determined by finding the best fitted line to the
data points at $<$3.26 $\mu$m and at 3.58--3.72 $\mu$m, as was done 
for NGC 3094.  
The adopted continuum level is shown in Fig. 2 as a solid
line.  The $\tau_{3.4}$ against this continuum level is $\sim$0.08.
Any other reasonable continuum levels provide consistent values 
for $\tau_{3.4}$ within 20\%. 

In this figure, the presence of the 3.3 $\mu$m PAH emission
feature is less clear than in the spectrum of NGC 3094.
At the redshift of {\it z} = 0.009, the wavelength of the peak of the
3.3 $\mu$m PAH emission feature ($\sim$3.29 $\mu$m in the rest-frame; 
Tokunaga et al. 1991) is redshifted to $\sim$3.32 $\mu$m, 
where the signals are 
extremely small due to the strong methane absorption by Earth's
atmosphere.  Hence, it is difficult to quantitatively estimate the 3.3
$\mu$m PAH emission flux in this spectrum.  
We tentatively estimate the rest-frame equivalent width of the 
3.3 $\mu$m PAH emission as $<$1.8 $\times$ 10$^{-3}$ $\mu$m.

It should be noted that Moorwood (1986) reported the detection of the
3.3 $\mu$m PAH emission feature with a 7$\farcs$5 aperture, although
the significance level of the detection was $<$3 $\sigma$.  
Moorwood's aperture size is larger than
that of Roche et al.  (1991; 4$\farcs$2) and of the present data
(1$\farcs$2 $\times$ 5$''$).  The radio map of this galaxy shows
extended emission in addition to unresolved emission (Unger et
al. 1987).  Since the direction of the extended radio emission is
roughly in the east-west direction, the same as that of the edge-on
host galaxy (Sharples et al. 1984; Anupama et al. 1995), 
the extended radio emission is
thought to originate from star-forming regions in the host galaxy.
These star-forming regions could explain the possible detection of the
3.3 $\mu$m PAH emission by Moorwood (1986) using a large aperture.

\subsection{NGC 7479}

A flux-calibrated spectrum of NGC 7479 is shown in Fig. 3.  
Our data give {\it L} $=$ 10.5 mag.  
Nearly all of the {\it L}-band flux within a 5$''$ aperture 
({\it L} = 10.4 mag; Willner et al. 1985) is detected in the 
spectrum (1$\farcs$2 $\times$ 5$''$ aperture).

An absorption-like feature at $\sim$3.35--3.55 $\mu$m is recognizable 
in the spectrum in Fig. 3.  
This feature is interpreted as the 3.4 $\mu$m carbonaceous dust 
absorption.
A continuum level is determined in the same way as was done for
NGC 3094 and NGC 7172.  
The adopted continuum level is shown in Fig. 3 as a solid
line.  The $\tau_{3.4}$ against this continuum level is $\sim$0.25.

The presence of the 3.3 $\mu$m PAH emission is not clear.
The rest-frame equivalent width of the 3.3 $\mu$m PAH emission 
is estimated to be $<$7.2 $\times$ 10$^{-3}$ $\mu$m.

\section{DISCUSSION} 

\subsection{Relation between the $\tau_{3.4}$/$\tau_{9.7}$ ratios 
and the inclination of the host galaxies}

In general, emission at $>$3 $\mu$m from the inner few arcsec of 
(moderately highly-luminous) obscured AGNs is  
dominated by a compact AGN component and not by extended stellar 
emission (Alonso-Herrero et al. 1998).
In this case, the observed optical depths of dust absorption features 
reflect the column density of dust in front of a background emission 
associated with AGN activity.

In the spectrum of NGC 3094, both the 3.4 $\mu$m carbonaceous 
dust absorption and the 3.3 $\mu$m PAH emission are detected.
Thus, its nuclear 3--4 $\mu$m emission is a composite of a 
highly obscured AGN and detectable star-forming activity.
In this case, the absorption-like feature at $\sim$10 $\mu$m is 
caused not only by the 9.7 $\mu$m silicate dust absorption but also 
by the 7.7 $\mu$m PAH emission.
The estimate of the column density of carbonaceous dust in front of 
a background obscured AGN based on the observed $\tau_{3.4}$ value 
is uncertain because star-forming activity contaminates the 3--4 $\mu$m 
continuum emission.

In the spectra of NGC 7172 and NGC 7479, 
the 3.4 $\mu$m carbonaceous dust absorption is detected, but 
no 3.3 $\mu$m PAH emission is detected.
Hence, their nuclear spectra are thought to be dominated by 
the emission of highly obscured AGNs.
In this case, the absorption-like feature at
$\sim$10 $\mu$m detected in ground-based 8--13 $\mu$m spectra is
ascribed mostly to the 9.7 $\mu$m silicate dust absorption.  
The ground-based 8--13 $\mu$m spectra of Roche et al. (1991) give the
optical depths $\tau_{9.7}$ of the 9.7 $\mu$m silicate dust absorption
feature as $>$2 and $\sim$2.5, respectively, for NGC 7172 and NGC 7479
(Table 2).  Using these $\tau_{9.7}$ values, the ratios
$\tau_{3.4}$/$\tau_{9.7}$ $<$ 0.04 and $\tau_{3.4}$/$\tau_{9.7}$
$\sim$ 0.10 are obtained towards the nuclei of 
NGC 7172 and NGC 7479, respectively.  
The $\tau_{3.4}$/$\tau_{9.7}$ ratios are 
0.06--0.07 towards any direction in the Galactic diffuse inter-stellar 
medium (Pendleton et al. 1994; Roche \& Aitken 1984, 1985).

In the literature, we find three more nuclei with detected dust 
absorption at 3.4 $\mu$m and 9.7 $\mu$m, and no detectable PAH 
emission features.
These three sources, together with NGC 7172 and NGC 7479, are summarized 
in Table 2. 
Among the five sources, NGC 7172 and NGC 5506 show
$\tau_{3.4}$/$\tau_{9.7}$ ratios smaller than the ratios in the 
Galactic diffuse inter-stellar medium, 
while NGC 7479, NGC 1068, and IRAS 08572+3915 show larger
$\tau_{3.4}$/$\tau_{9.7}$ ratios.

Since the Galactic extinction is negligible towards all five nuclei 
($A_{\rm V}$ $<$ 1 mag; Burstein \& Heiles 1984; 
the Einstein On-Line Service
\footnote{The ftp address is 131.142.11.73.}
), 
the dust extinction must be outside the Galaxy.
The host galaxies of NGC 7172 and NGC 5506, the two sources with 
smaller $\tau_{3.4}$/$\tau_{9.7}$ ratios, are seen from 
an almost edge-on direction 
(Sharples et al. 1984; Anupama et al. 1995; Whittle 1992), 
while the viewing directions towards the host galaxies of 
NGC 7479, NGC 1068, and IRAS 08572+3915, the three sources with 
larger $\tau_{3.4}$/$\tau_{9.7}$ ratios, are relatively face-on 
(Ma, Peng \& Gu 1998; Whittle 1992; Surace et al. 1998). 
A trend is found, that the $\tau_{3.4}$/$\tau_{9.7}$ ratios are smaller 
than the Galactic value in two nuclei with edge-on host galaxies, 
while the ratios are larger in three nuclei with face-on host galaxies.

We interpret all five nuclei as powered predominantly by 
obscured AGN activity, based on the absence of detectable 
PAH emission features at 3--4 $\mu$m and 8--13 $\mu$m.
However, since the detection feasibility of the PAH emission depends on 
the spectral quality of individual sources, 
emission of star-forming activity could be non-negligible in some 
of the five nuclei.
In this case, the $\tau_{3.4}$ value could be suppressed 
and the apparent $\tau_{9.7}$ value could increase (section 1), 
that is, the $\tau_{3.4}$/$\tau_{9.7}$ ratio could decrease.
Hence, although the interpretation of smaller $\tau_{3.4}$/$\tau_{9.7}$ 
ratios requires some cautions, the larger $\tau_{3.4}$/$\tau_{9.7}$ 
ratios cannot be caused by the possible contamination of the emission 
of star-forming activity.

\subsection{Possible explanations for the smaller and larger  
$\tau_{3.4}$/$\tau_{9.7}$ ratios}

First, smaller $\tau_{3.4}$/$\tau_{9.7}$ ratios are found towards 
two highly obscured AGNs whose host galaxies are seen from an 
edge-on direction. 
It seems reasonable that dust in the host galaxies are responsible 
for the obscuration 
(Young et al. 1996; Murayama, Mouri \& Taniguchi 2000).
The galaxy types of these host galaxies are S0--Sa 
(Anupama et al. 1995; Kinney et al. 1991), earlier than that of 
the Galaxy (Sb/bc; Kerr 1993).
If the contribution by carbonaceous dust to dust extinction is smaller 
in these host galaxies, the smaller $\tau_{3.4}$/$\tau_{9.7}$ ratios 
could be explained.

Next, larger $\tau_{3.4}$/$\tau_{9.7}$ ratios are found in 
three highly obscured AGNs for which the viewing directions towards 
their host galaxies are relatively face-on.
Carbonaceous dust is more easily destroyed than silicate dust 
(Draine \& Salpeter 1979). 
If the obscuring dust towards the three AGNs suffers less dust 
destruction than the Galactic diffuse inter-stellar medium, then 
the larger $\tau_{3.4}$/$\tau_{9.7}$ ratios could be explained.
However, we suggest that the larger ratios can be explained more 
naturally by the presence of a temperature gradient in the obscuring 
dust.
If the obscuring dust is located close to a central compact energy 
source, a temperature gradient is predicted to occur, 
with the temperature of the dust decreasing with increasing distance 
from the central engine (Pier \& Krolik 1992).  
The temperature of the innermost dust ($<$ a few pc) is expected 
to be $\sim$1000 K, close to the dust sublimation temperature.  
Since emission at $\sim$3 $\mu$m is dominated by dust at $\sim$1000 K, 
the extinction estimated using the $\sim$3 $\mu$m data 
(i.e., observed $\tau_{3.4}$) should reflect the value towards the 
innermost dust around the central energy source.
On the other hand, dust at $\sim$300 K, 
a dominant emission source at 10 $\mu$m, is located further out 
than the $\sim$1000 K dust, and thus the extinction estimated using 
the $\sim$10 $\mu$m data (i.e., observed $\tau_{9.7}$) is only 
towards the outer region.
Hence, the $\tau_{3.4}$/$\tau_{9.7}$ ratios can be significantly larger.
If this suggestion is correct, our results may provide 
spectroscopic evidence for the presence of the putative ``dusty tori'' 
in the close vicinity ($<$ a few pc in inner radius) of AGNs 
(Antonucci 1993).

\section{SUMMARY}  

The following main results have been found.

\begin{enumerate}
\item The 3.4 $\mu$m carbonaceous dust absorption feature has been
detected towards the nuclei of NGC 3094, NGC 7172, and NGC 7479.
These nuclei are those with a strong absorption-like feature 
at $\sim$10 $\mu$m but with no detectable PAH emission features 
in ground-based 8--13 $\mu$m spectra.
NGC 3094 shows a detectable 3.3 $\mu$m PAH emission feature, while 
NGC 7172 and NGC 7479 do not.
The detection of the 3.4 $\mu$m absorption, together with the
non-detection of the 3.3 $\mu$m PAH emission, suggests that the 
nuclei of NGC 7172 and NGC 7479 are powered by 
highly obscured AGN activity. 
Their strong absorption-like feature at $\sim$10 $\mu$m should be 
attributed mostly to 9.7 $\mu$m silicate dust absorption.
On the other hand, the detection of both 3.4 $\mu$m absorption and 
3.3 $\mu$m PAH emission in NGC 3094 suggests that the 
nucleus is powered both by highly obscured AGN activity and detectable 
star-forming activity.
Both the 9.7 $\mu$m silicate dust absorption and the 7.7 $\mu$m PAH 
emission could be responsible for the strong absorption-like feature 
at $\sim$10 $\mu$m.

\item We compared the $\tau_{3.4}$/$\tau_{9.7}$ ratios for five 
highly obscured AGNs with detected 3.4 $\mu$m and 9.7 $\mu$m 
absorption and without detectable PAH emission features.
We found that the ratios in the three highly obscured AGNs whose 
host galaxies are seen from a face-on direction are larger than 
the ratios in the Galactic diffuse inter-stellar medium or the ratios 
in the two highly obscured AGNs whose host galaxies are seen 
from an edge-on direction.
For the three sources, the larger ratios can be explained if the 
obscuring dust is so close to the central AGNs that a temperature 
gradient occurs in it. 
Our results may provide spectroscopic evidence for the presence of 
the putative ``dusty tori'' around AGNs.

\end{enumerate}

\section*{Acknowledgments}      

We thank Dr. T. Kerr, Dr. J. Davies, T. Carroll, and T. Wold for their 
support during the UKIRT observing run, and Dr. C. C. Dudley for his 
careful reading of the manuscript.
L. Good kindly proofread this manuscript.
The anonymous referee gave useful comments.
The United Kingdom Infrared Telescope is operated by the Joint Astronomy 
Centre on behalf of the U.K. Particle Physics and Astronomy Research 
Council.
Drs. A. T. Tokunaga and H. Ando give MI the opportunity to work at 
the University of Hawaii.
MI is financially supported by the Japan Society for the Promotion 
of Science during his stay at the University of Hawaii.

\clearpage

\small
\begin{table*}
\begin{center}

\caption{Observing log}
\begin{tabular}{ccccccccc} \hline \hline
Object & $z$ & Date & Integration & Air & \multicolumn{4}{c}
{Standard Star} \\ 
 & & (UT) & Time (sec) & Mass & Name & $L$-mag & Type & 
Temperature (K) \\ \hline
NGC 3094 & 0.008 & Feb 21, 2000 & 2560 & 1.0--1.2 & HR 3975 & 3.3 & A0Ib 
& 9700 \\
NGC 7172 & 0.009 & Sep 9, 1999  & 3200 & 1.6--1.7 & HR 8087 & 5.3 & A0V  
& 9480 \\
NGC 7479 & 0.008 & Sep 9, 1999  & 2400 & 1.0--1.1 & HR 8911 & 4.9 & A0 
& 9480 \\
\hline
\end{tabular}
\end{center}
\end{table*}

\normalsize



\begin{table*}
\begin{center}

\caption{The $\tau_{3.4}$, $\tau_{9.7}$ and $\tau_{3.4}$/$\tau_{9.7}$ 
towards sources with detected 3.4 $\mu$m and 9.7 $\mu$m absorption features  
and with no detectable PAH emission features.}
\begin{tabular}{llcccc} \hline \hline
Object & $z$ & $\tau_{3.4}$ & $\tau_{9.7}$ & $\tau_{3.4}$/$\tau_{9.7}$ 
& References \\ \hline
NGC 1068 & 0.004 & 0.12 & 0.52 & 0.23 & 1, 2 \\
IRAS 08572+3915 & 0.058 & 0.9  & 5.2 & 0.17 & 3, 4 \\
NGC 7479 & 0.008 & 0.25 & $\sim$2.5 & 0.10& 5, 6 \\
NGC 5506 & 0.006 & 0.03 & 1.3 & 0.02 & 7, 2 \\
NGC 7172 & 0.009 & 0.08 & $>$2 & $<$0.04 & 5, 6 \\
\hline
\end{tabular}

References. --- (1) Imanishi et al. 1997; 
(2) Roche et al. 1984; 
(3) Pendleton 1996; 
(4) Dudley \& Wynn-Williams 1997; 
(5) this work; 
(6) estimated based on the spectra of Roche et al. (1991). 
    We estimate $\tau_{9.7}$ by (a) using the formula of 
    $\tau_{9.7}$ $=$ ln [$\frac{F_{\rm \lambda}(8) + F_{\rm \lambda}(13)}
    {2F_{\rm \lambda}(9.7)}$] (Aitken \& Jones 1973), and by (b) 
    deriving the optical depth at 9 $\mu$m ($\tau_{9}$) and by 
    adopting the relation of $\tau_{9.7}$/$\tau_{9}$ $\sim$ 1.7  
    (Dudley \& Wynn-Williams 1997).
    Both methods provide consistent results;   
(7) Imanishi 2000. 

\end{center}
\end{table*}

\clearpage

\begin{figure*}
\centerline{\psfig{file=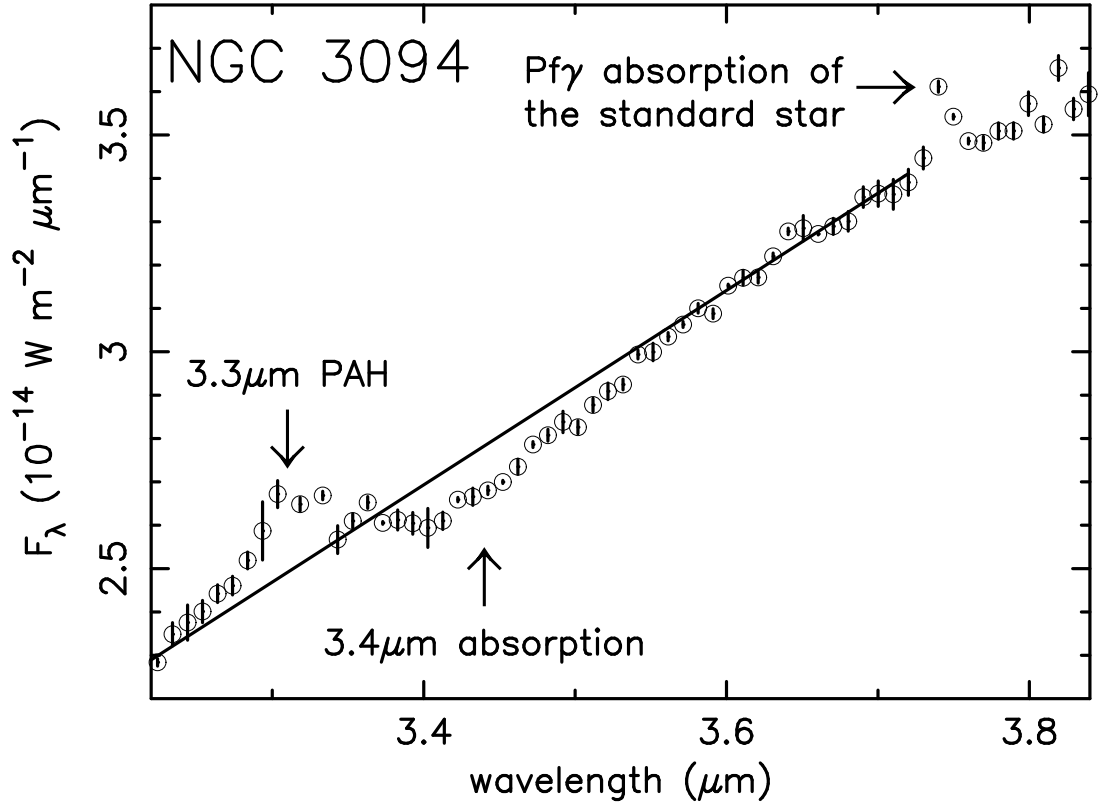,angle=0,width=6.5in}}
\caption{
A flux-calibrated spectrum of NGC 3094.
A spectral resolution is $\sim$350 after binning.
The ordinate is F$_{\lambda}$ in W m$^{-2}$ $\mu$m$^{-1}$, and the 
abscissa is the observed wavelength in $\mu$m.
The solid line is the adopted continuum level to measure the 
optical depth of the 3.4 $\mu$m absorption feature (see text).
The emission-like feature at 3.74 $\mu$m is caused by 
the stellar absorption of the type-A standard star (see text).
}
\end{figure*} 

\begin{figure*}
\centerline{\psfig{file=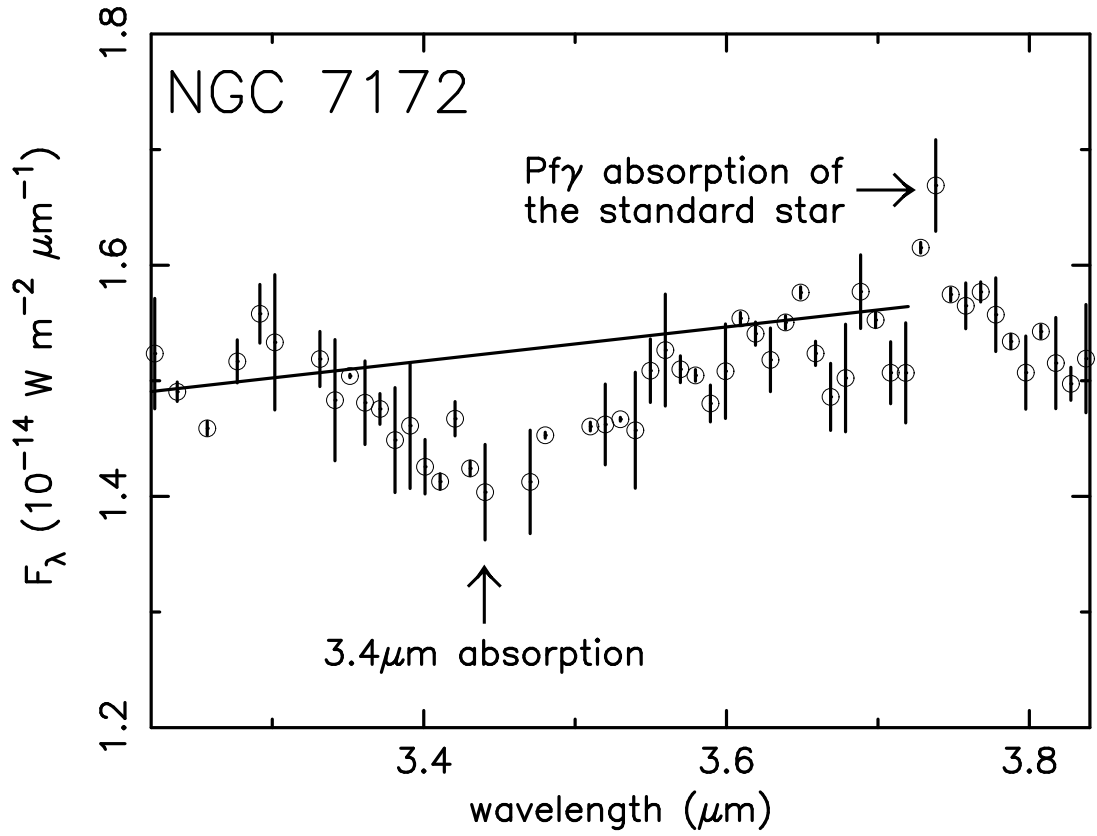,angle=0,width=6.5in}}
\caption{
A flux-calibrated spectrum of NGC 7172, displayed in the same way as 
NGC 3094 (Fig. 1).
}
\end{figure*} 

\newpage

\begin{figure*}
\centerline{\psfig{file=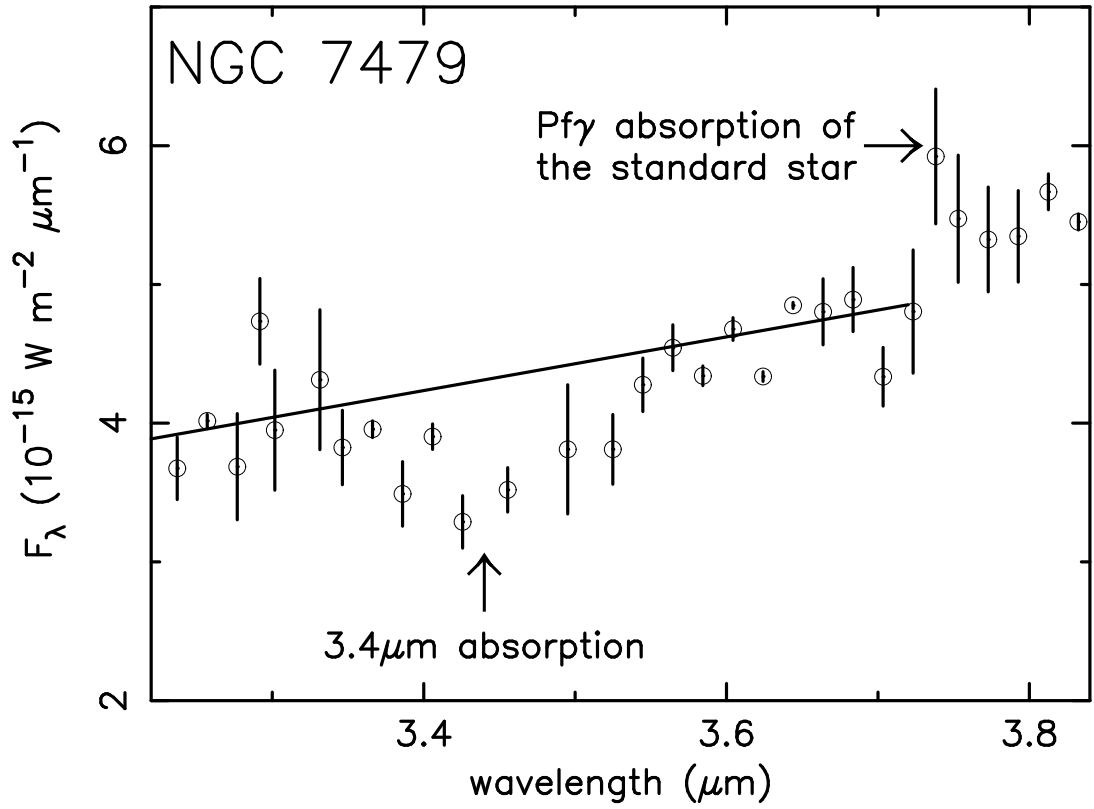,angle=0,width=6.5in}}
\caption{
A flux-calibrated spectrum of NGC 7479, displayed in the same way as 
NGC 3094 (Fig. 1).
This spectrum is shown with a spectral resolution of $\sim$170 
at 3.5 $\mu$m. 
}
\end{figure*} 


\end{document}